\documentclass[prl,twocolumn,showpacs,amsmath,amssymb]{revtex4-1}
\usepackage{graphicx}
\usepackage{dcolumn}
\usepackage{bm}
\begin{document}
\title{Modeling flux noise in SQUIDs due to hyperfine interactions} 
\author{Jiansheng Wu and Clare C. Yu}
\affiliation{Department of Physics and Astronomy, University of California, 
Irvine, California 92697-4575}
\date{\today}
\begin{abstract}
Recent experiments implicate spins on the surface of metals as the source
of flux noise in SQUIDs, and indicate that these spins are able to relax
without conserving total magnetization. We present a model of $1/f$ flux
noise in which electron spins on the surface of metals can relax via hyperfine 
interactions. Our results indicate that
flux noise would be significantly reduced in superconducting materials
where the most abundant isotopes do not have nuclear moments such as zinc and lead.
\end{abstract}

\maketitle

Although there have been significant advances in superconducting qubits, they
continue to be plagued by noise and decoherence. 
Low frequency $1/f$ flux noise \cite{Wellstood1987} in superconducting 
quantum interference devices (SQUIDs) is one of the dominant sources of noise in
superconducting flux \cite{Yoshihara2006,Kakuyanagi2007}
and phase \cite{Bialczak2007} qubits. Recent experiments indicate
this flux noise arises from the fluctuations of spins residing on
the surface of normal metals \cite{Bluhm2009} and 
superconductors \cite{Sendelbach2008}.
These spins have a high density ($\sim 5\times 10^{17}$ m$^{-2}$), and
may arise from local electron moments in localized states
at the metal-insulator interface \cite{Choi2009}. 

One early model of flux noise due to spins proposed that the spin
of an electron in a surface trap is fixed, but that the orientation of
the spin can
change when the electron hops to a different trap \cite{Koch2007}. 
However, the density of defect traps needed to explain the experiments
was orders of magnitude larger than what is estimated to exist
in a typical glassy material \cite{Bialczak2007}. Another model
suggested that spin flips of paramagnetic dangling bonds occurred
as a result of interactions with tunneling two-level systems mediated
by phonons \cite{Sousa2007}. However, to obtain $1/f$ flux noise,
the maximum two level system energy splitting would have to be a few
mK which is orders of magnitude smaller than accepted values.

There is some experimental indication
of interactions between the spins \cite{Sendelbach2008} leading to the
theoretical suggestion that flux noise is the result of 
spin diffusion via
Ruderman-Kittel-Kasuya-Yosida (RKKY) interactions \cite{RKKY} 
between the spins \cite{Faoro2008}.
RKKY interactions between randomly placed spins produce spin glasses, and
Monte Carlo simulations of Ising spin glass systems show that interacting 
spins produce 1/f flux and inductance noise \cite{Chen2010}.

In addition, RKKY interactions conserve the total spin and magnetization, 
so the total magnetization should not change. However, the Stanford group 
measured the total magnetization of small isolated gold
rings and found that the total magnetization is not conserved since the
magnetization follows the externally applied ac magnetic field \cite{Bluhm2009}.
While this does not rule out magnetization conserving interactions such as RKKY, it
does imply that the spins must (also) be involved in interactions that do not conserve
total magnetization. (The angular momenta contained in the external magnetic field 
and in the electric current
induced in the gold ring are orders of magnitude too small to conserve
total angular momentum by accomodating the change
in angular momentum of the spins associated with the change in total magnetization.) 

The 3 possible interactions that do not conserve total magnetization
are spin-orbit, magnetic dipole-dipole, and hyperfine
interactions. The orbital angular momentum of a neutral gold atom is zero, so we
can ignore spin-orbit interactions. The dipole-dipole interaction between 2 
electrons is of order 1 mK if we use a separation of 1.4 nm corresponding to 
a spin density of $5\times 10^{17}$ m$^{-2}$. This is much smaller than
the hyperfine contact interaction which is of order 70 mK in hydrogen, for example. 
This implies that
the hyperfine interaction dominates. In support of this is the fact that
scanning SQUID microscope experiments \cite{Bluhm2009} found that the 
magnetic susceptibility of spins on silicon is 5 to 20 times smaller 
than that of metals and insulating AlO$_x$. This is consistent with 
hyperfine interactions since the only isotope of
silicon that has a nuclear spin and hence hyperfine interactions is $^{29}$Si 
which has a natural isotopic abundance of 5\%.
Note that spin angular momentum is conserved in hyperfine interactions where 
there is a spin flip exchange between the nuclear spin and the electron spin, 
but the magnetic moment is not conserved since the gyromagnetic ratios of the 
nuclear and electron spins differ by about 3 orders of magnitude.
Previous authors \cite{Koch2007,Faoro2008} have pointed out
that flux noise cannot be directly due to
fluctuating nuclear spins because the frequency range and magnitude of nuclear
flux noise would be much lower than what is seen experimentally. 
However, this does not rule out the possibility that the
electron spins that are responsible for flux noise
can relax via hyperfine interactions with nearby nuclear spins.

In this paper we present a model of flux noise in which electrons residing in 
harmonic traps undergo spin exchange with nearby nuclei via the hyperfine contact
interaction. The relaxation time $T_1$ of a given electron spin is dominated
by exchange with the nearest nonzero nuclear moment. In materials where
not all the isotopes have a nuclear moment, the distance to the nearest
nucleus with a nuclear spin could be quite large. For example, the only 
isotope of Pb with a nuclear moment is $^{209}$Pb which has a 22\% natural 
isotopic abundance. The distribution of distances between trapped electrons
and the nearest nucleus with a magnetic moment gives rise to a distribution 
of electron spin relaxation times $T_1$, which in turn results in 1/f noise
up to 10 MHz. Unlike the model \cite{Faoro2008} of spin diffusion via RKKY 
that found white noise at low frequencies in contradiction to experiment, 
we find that 1/f flux noise extends down to $10^{-5}$ Hz.

The Hamiltonian of an electron spin $\bf S$ that is in an
external field $\bf H_{\rm ext}$ and that has a contact hyperfine coupling
to nearby nuclear spins ${\bf I}_i$ is given by \cite{Abragam1961}
\begin{eqnarray}
  {\cal H} &=& {\cal H}_0+ {\cal H}_{hyp} \nonumber\\
  {\cal H}_0 &=& -g\mu_B\ {\bf H}_{\rm ext}\cdot {\bf S} \nonumber\\
             &\equiv& -\hbar\omega_0 S_z \nonumber\\
   {\cal H}_{hyp} &=& \sum_i \frac{8\pi}{3} \frac{\mu_0}{4\pi} g_0 \mu_B \gamma_n \hbar
                      {\bf I}_i\cdot {\bf S}\delta({\bf r-r_i}),
\end{eqnarray}
where $I_z=\pm 1$ and $S_z=\pm 1$.
We ignore the dipolar interaction between the electron and the nuclear spins 
because it is much smaller than the contact hyperfine interaction \cite{Lu2006}.
We choose ${\bf H}_{\rm ext}$ parallel to ${\hat z}$. 
$\mu_0$ is the permeability constant.
The external field could be due to an applied external
magnetic field or to the magnetic field produced by local electric currents.
${\bf r}_i$ and ${\bf r}$ are the coordinates of the i-th nuclei and the electron, 
$g_0$ is the free-electron g-factor and $\gamma_n$ is the nuclear gyromagnetic ratio.
$\mu_B$ is the Bohr magneton of the electron.
Taking the expectation value of ${\cal H}_{hyp}$ with respect to the 
electron wavefunction $\psi(r)$ yields
\begin{equation}
\langle {\cal H}_{\rm hyp}\rangle_e=\frac{2}{3}\mu_0 g_0\mu_B \gamma_n \hbar
\sum_{i} {\bf I}_i \cdot {\bf S} |\psi({\bf r_i})|^2,
\end{equation}
where $\psi({\bf r_i})$ is the wavefunction of the electron at the position 
of the $i$th nucleus. $\langle \rangle_{e}$ indicates the expectation value 
of the Hamiltonian ${\cal H}_{\rm hyp}$ with respect to the electron wavefunction.
This can be written in the standard form of a hyperfine interaction:
\begin{equation}
\langle {\cal H}_{\rm hyp}\rangle_{e} =\sum_i ( A_{\rm hf}^i\ \bf I_i)\cdot {\bf S}
\end{equation}
$A_{\rm hf}^i$ is the hyperfine coupling constant between the $i$th nuclear spin 
and the electron spin. A typical hyperfine frequency, e.g., for hydrogen, is
$f_{hf}=A_{\rm hf}/h\sim 1.4$ GHz where $h$ is Planck's constant.
We can also express this in terms of an effective random field ${\bf H}_{\rm R}^i$
produced by the $i$th nuclear spin:
\begin{equation}
\langle {\cal H}_{\rm hyp}\rangle_{e} = \sum_i g_0\mu_B{\bf H}_{\rm R}^i \cdot {\bf S}
\end{equation}
where
\begin{equation}
{\bf H}_{\rm R}^i = \frac{2}{3}\mu_0 \gamma_n \hbar |\psi({\bf r_i})|^2
{\bf I_i}
\label{eq:HR}
\end{equation}

To find the effective field ${\bf H}^i_{\rm R}$ on the electron due to the 
$i$th nucleus, we use Eq.~(\ref{eq:HR}) and assume 
that the electron is in a harmonic trap with a ground state wavefunction
\begin{equation}
\psi(r)=\frac{1}{\sqrt{\pi\xi^2}}e^{-r^2/2\xi^2}
\end{equation}
where $\xi^2=\hbar/m_e\omega$, $m_e$ is electron mass and $\Omega$ is the 
frequency of the harmonic oscillator. 
(If we assume that the localized wavefunction decays exponentially as 
$\exp(-r/\xi_{\ell})$
where $\xi_{\ell}$ is the localization length, then we still obtain $1/f$ noise
up to logarithmic corrections in the frequency.)
This gives
\begin{eqnarray}
{\bf H}_{\rm R}^i &=&\frac{2}{3}\frac{\mu_0  \gamma_n \hbar}{\pi\xi^2}
{\bf I}_i e^{-r^2_{i}/\xi^2}\nonumber\\
&\equiv &\frac{A_0}{\xi^2}{\bf I}_{i} e^{-r^2_{i}/\xi^2}
\label{eq:HR_SHO}
\end{eqnarray}
where $A_0$ is a constant.

The nuclear spin dynamics can be characterized by a correlation time $\tau_0$
that is roughly the time scale over which the nuclear spins keep their orientation
\cite{Slichter1981}
\begin{eqnarray}
C_j^i(t)&\equiv &\left< H_{\rm R,j}^i(t+\tau)H^i_{\rm R,j}(t)\right>\nonumber\\
&=&\left<(H^i_{\rm R,j})^2\right>\exp(-|\tau|/\tau_0),
\end{eqnarray}
where $j=x,y,z$ are the components of the random magnetic fields.
The Fourier transform of this correlation function is
\begin{eqnarray}
C_j^i(f)&=& \left<(H^i_{\rm R,j})^2\right>\frac{\tau_0}{1+(2\pi f)^2\tau_0^2} , 
\label{noise}
\end{eqnarray}
We will regard $\tau_0$ as a constant. If $\tau_0$ is determined by 
spin diffusion via
nuclear dipole-dipole interactions, then $\tau_0\sim 1/Dq^2$ where $D$ is the 
spin diffusion constant and $q=\pi/a$ where $a$ is the typical distance between
nuclear spins.

The electron spin dynamics is given by the Bloch equation \cite{Slichter1981, Sousa2006}:
\begin{equation}
\frac{d}{dt}\left< {\bf S}\right> =- \mu_B\ {\bf H}_{\rm ext}\times 
\left< {\bf S}\right>- \frac{1}{T_1} \left< S_z \right>  {\hat{z}}-
\frac{1}{T_{2x}} \left< S_x\right>  {\hat{x}} -
\frac{1}{T_{2y}}\left< S_y\right> {\hat{y}}, 
\label{meqn}
\end{equation}
where $T_1$ is the spin-lattice relaxation time and $T_2$ is the spin-spin 
relaxation time defined by \cite{Slichter1981, Sousa2006}
\begin{eqnarray}
\frac{1}{T_1}&=&
\left(\left< (H^i_{\rm R,x})^2\right>+\left<(H^i_{\rm R,y})^2\right>\right)
\frac{\tau_0}{1+\omega_0^2\tau_0^2},
\label{T1}\\
\frac{1}{T_{2x}}&=& 
\left<(H^i_{\rm R,z})^2\right>\tau_0 
+\left<(H^i_{\rm R,y})^2\right>\frac{\tau_0}{1+\omega_0^2\tau_0^2},
\label{T2x}\\
\frac{1}{T_{2y}}&=& 
\left<(H^i_{\rm R,z})^2\right>\tau_0 +\left<(H^i_{\rm R,x})^2\right>
\frac{\tau_0}{1+\omega_0^2\tau_0^2}.
\label{T2y}
\end{eqnarray}
Here we used Eq.~(\ref{noise}).

Each nuclear spin produces an electron relaxation time $T_1({\bf r}_i)$ that
we can obtain by plugging Eq.~(\ref{eq:HR_SHO}) into Eq.~(\ref{T1}) to yield:
\begin{eqnarray}
\frac{1}{T_1({\bf r}_i)}&=&\frac{C^2\tau_0}{{\xi^4}(1+\omega_0^2\tau_0^2)}
e^{-2 r_i^2/\xi^2}\nonumber\\
&=& \frac{b}{\xi^4}e^{-2r^2_{i}/\xi^2}
\end{eqnarray}
where $C$ and $b$ are constants.
To find an effective relaxation rate $1/T_1^{\rm eff}$, we sum or integrate over
all the nuclei that lie within the electron wavefunction. The relaxation
rate will be dominated by the nuclear spin that is closest to the electron. 
Let $r_n$ denote the distance between the center of the electron wavefunction 
and nuclear spin closest to it. Then we obtain
\begin{eqnarray}
\frac{1}{T_1^{\rm eff}}&=&\sum_{i}\frac{1}{T_1({\bf r}_i)}\nonumber\\
&\rightarrow &\frac{1}{a^2}\int_{r_n}^{\infty} d^2 {\bf r} 
\frac{1}{T_1({\bf r})}\nonumber\\
&=&\frac{\pi b}{2\xi^2 a^2}e^{-2 r_n^2 /\xi^2},
\end{eqnarray}
where $a$ is the lattice constant.
We now need to average over all the localized electron spins on the surface. 
Since the electrons are uniformly and randomly distributed on the surface, 
$r_n$ has a distribution $P(r_n)$.
For a 2D square lattice, the distribution of $r_n$ is $P(r_n)=2\pi r_n/a^2 $ for 
$0\le r_n\le a/2$. The surface on which the spins sit can be 
disordered, possibly resulting in an exponent for $r_n$ that is different from unity. 
However, it is reasonable to assume $P(r_n)=A_{1}r_n^{\gamma-1}/a^{\gamma}$ 
where $A_1$ is a constant and $\gamma\in(2,4)$ ($\gamma$ is of order the dimension).
Since $r_n$ depends logarithmically on $T_1^{\rm eff}$
for both an electron wavefunction in a harmonic trap and an exponentially localized
electron, the actual form of $P(r_n)$ is not important. The resulting 
distribution of $T_1^{\rm eff}$ is
\begin{equation}
P(T_1^{\rm eff})=\frac{A_1}{2^{\gamma/2+1}}\left(\frac{\xi}{a}\right)^{\gamma}
\left[\ln\left(\frac{\pi bT_1^{\rm eff }}{2\xi^2 a^2}\right)\right]^{\frac{\gamma-2}{2}}
\frac{1}{T_1^{\rm eff}}.
\label{eq:PT1}
\end{equation}
We can simplify the above formula by approximating the slowly varying function 
$\ln\left(\pi bT_1^{\rm eff} /2\xi^2 a^2\right)$ by its average value. 
Then the distribution function is inversely proportional to $T_1^{\rm eff}$,
and we can write $P(T_1^{\rm eff})=D_1/T_1^{\rm eff}$ where $D_1$ is a normalization
factor determined by
\begin{equation}
\int_{T_{1,min}^{\rm eff}}^{T_{1,max}^{\rm eff}}dT_1^{\rm eff}P(T_1^{\rm eff})=1
\end{equation}
$\left(T_{1,\rm max}^{\rm eff}\right)^{-1}$ and 
$\left(T_{1,\rm min}^{\rm eff}\right)^{-1}$ correspond to the minimum and maximum
frequencies of the flux noise, and are determined by
$(r_n)_{\rm max}$ and $(r_n)_{\rm min}$, respectively. Thus we find
$D_1=\xi^2/2\left[(r^2_n)_{\rm max}-(r^2_n)_{\rm min}\right]\equiv \xi^2/2\Delta r_n^2$.

According to the Wiener-Khintchine theorem, the spectral density $S(\omega)$ 
of the noise is given by twice the Fourier transform
of the autocorrelation function of the spin fluctuations.
From the fluctuation-dissipation theorem, the low frequency 
($\hbar\omega\ll kT$) spin noise is proportional
to the imaginary part of the spin susceptibility that we can derive from
the Bloch equations in Eq.~(\ref{meqn}) \cite{Slichter1981}. The frequency
dependence of the noise at low frequencies is determined by the z-component
of the susceptibility \cite{Shnirman2005}. The resulting spin noise power
is \cite{Dutta1981}
\begin{eqnarray}
\mathcal{S}_z(\omega) &=& 
2\int^{T^{\rm eff}_{1,\rm max}}_{T^{\rm eff}_{1,\rm min}} d{T_1^{\rm eff}}P(T_1^{\rm eff}) 
{\rm sech}^2\left(\frac{\hbar\omega_0}{k_B T}\right)
\frac{1/T_1^{\rm eff} }{\omega^2+(1/T_1^{\rm eff})^2}\nonumber \\
    &\approx&  \frac{\xi^2}{\Delta r_n^2} {\rm sech}^2
\left(\frac{\hbar\omega_0 }{k_B T}\right)\frac{\pi}{\omega},
\end{eqnarray}
where the limits of integration have a wide range with
$\omega T^{\rm eff}_{1,\rm min}\ll 1\ll \omega T^{\rm eff}_{1,\rm max}$.
 
To relate $\mathcal{S}_z(\omega)$ to the flux noise, we need to know how a 
spin couples magnetically to the SQUID. The effective
flux $\Phi_{\rm eff}$ produced by the spin magnetization on a loop with current
$I$ is \cite{Faoro2008}
\begin{equation}
    \Phi_{\rm eff}=g\mu_B\int \frac{{\hat{S}}({\bf r}) B({\bf r})}{I}d {\bf r},
\label{flux}
\end{equation}
where ${\hat{S}}({\bf r})$ is the surface spin density operator and $B(r)$ denotes
the probing magnetic field due to the current and, if applicable, an externally
applied field. Consider a SQUID made from a strip conductor (where the
width of the strip is $d$) circular in shape with
radius $R$ (measured from the center of the loop to the middle of the annulus).
(For a square SQUID with circumference $L$ and width $W$, we replace $R$
and $d$ by $L/2\pi$ and $W$, respectively.) If the penetration depth $\lambda$
is much smaller than the width, the current density at $x$ near the center of the
strip is $J(x)=2I/(\pi d)[1-(2x/d)^2]^{-1/2}$ for
$(-d/2)+\lambda<x<(d/2)-\lambda$ \cite{McDermott2008,Faoro2008}. This
current density produces the magnetic field 
$B(x)=\mu_0 J(x)/2$. 
Using this in Eq.~(\ref{flux}), we obtain the
flux autocorrelation function \cite{Faoro2008, McDermott2002}:
\begin{equation}
  \left<\Phi(t)\Phi(0)\right> = \frac{(g\mu_B)^2 R}{I^2}
\int_{-\frac{d}{2}}^{\frac{d}{2}}d{\bf r }d{\bf r'}\left<\hat{S}({\bf r},t)B({\bf r})
  \hat{S} ({\bf r'},0)B({\bf r'})\right> 
\label{eq:SSBB}
\end{equation}
If we assume the spins are isolated,
\begin{eqnarray*}
\langle{\hat{S}}({\bf r},t){\hat{S}}({\bf r'},0)\rangle &=&
\Theta(\sqrt{A}-|{\bf r-r'}|)\langle\hat{ S} ({\bf r},t)\hat{S} ({\bf r},0)\rangle\\
&=& \sigma^2 \mathcal{S}_z (t) \Theta(\sqrt{A}-|{\bf r-r'}|)
\end{eqnarray*} 
where $\sigma=1/A$ is the spin surface density, $A$ is the average area per spin,
$\Theta(x)$ is a step-function, and $\mathcal{S}_z (t)$ is the spin 
fluctuation autocorrelation function. 
After integrating over ${\bf r'}$ (using 
$\int \Theta(\sqrt{A}-|{\bf r-r'}|)f({\bf r'})d{\bf r'}\approx A f({\bf r})$ 
for an arbitrary function $f({\bf r})$), we obtain
\begin{equation}
  \left<\Phi(t)\Phi(0)\right> = \sigma \frac{(g\mu_B\mu_0)^2}{\pi}
\frac{R}{d}\ln\left(\frac{d}{2\lambda}\right)\mathcal{S}_z (t)
\label{FluxSpectrum}
\end{equation}
The associated flux noise spectrum is
\begin{equation}
  \mathcal{S}_{\Phi}(f)  = \sigma {(g\mu_B\mu_0)^2}{\rm sech}^2 
\left(\frac{\hbar\omega_0}{k_B T}\right) 
  \frac{R}{d}\ln\left(\frac{d}{2\lambda}\right) \frac{ \xi^2}{\Delta r_n^2}\frac{1}{2\pi f},
\label{eq:FluxS}
\end{equation}
which give rise to $1/f$ flux noise. Notice that the flux noise is proportional to
the density $\sigma$ of electron spins. Note also that for materials with a low
concentration of nuclei with magnetic moments, $\Delta r_n^2$ will be larger and
the flux noise, which goes as $1/\Delta r_n^2$, will be smaller. This is consistent
with the small susceptibility found on silicon samples \cite{Bluhm2009} where
$^{29}$Si is the only isotope with a nuclear moment and its natural isotopic
abundance is only 5\%. 

Let us estimate the flux noise magnitude at 1 Hz for a Josephson junction. We can
set the temperature factor to unity since 
$(\mu_BH_{\rm ext}=\hbar\omega_0)\ll k_B T$ 
for $H_{\rm ext}$ in the range of 1 to 100 G and $T$ between 25 mK and 10 K.
Since $(r_n)_{\rm min}\approx 0$ and we estimate that
$(r_n)^2_{\rm max}/\xi^2 > 30$,
we make the approximation that 
$\xi^2/\Delta r_n^2=\xi^2/2\left[(r^2_n)_{\rm max}-(r^2_n)_{\rm min}\right]\approx 1/30$.
Using $R/d=10$, $\sigma=5 \times 10^{17}\rm m^{-2}$ \cite{Sendelbach2008}, 
$\ln(d/2\lambda)\sim 8.5$, and 
$g\mu_0\mu_B\sim 11.3(\mu\Phi_0)$(nm), 
we estimate the amplitude of the flux 
noise to be $\mathcal{S}_{\Phi,\rm hf}^{1/2}\approx 5\ \rm \mu\Phi_0/\rm Hz^{1/2}$. 
This agrees with experimental values which are typically in the range of
1 to 10 $\mu\Phi_0/\rm Hz^{1/2}$ \cite{Yoshihara2006,Wellstood1987a}.

Eq.~(\ref{eq:FluxS}) gives the flux noise due to spins that are only on 
the surface of SQUIDs. However, paramagnetic spins have also been found 
on a diectric surface \cite{Bluhm2009}. So
if unpaired spins also reside on the substrate, these fluctuating spins will
also contribute to the flux noise, reducing the dependence on $d$.
Let $L$ be the self-inductance of the SQUID. Then
we can follow Wellstood \cite{Wellstood2009} and
use the expression for the electromagnetic energy
$E=LI^2/2=\int |B(r)|^2 d^2{\bf r}/(2\mu_0)$ to evaluate the integral in
Eq. (\ref{eq:SSBB}): 
\begin{equation}
  \mathcal{S}_{\Phi}(f)  = \sigma {\mu_0(g\mu_B)^2}{\rm sech}^2 
\left(\frac{\hbar\omega_0}{k_B T}\right) 
 {\it L} \frac{ \xi^2}{\Delta r_n^2}\frac{1}{2\pi f},
  \label{eq:FluxSL}
\end{equation}
The dependence of the flux noise on the geometry and the substrate are included
in $L$. This result agrees with recent experiments \cite{Sank2011} that found
a nearly linear relationship between flux noise and $L$ when the inductance of the
SQUID was enhanced by inductor coils.
Furthermore, our result also implies that flux noise and inductance noise should be
correlated \cite{Sendelbach2009}.  

To summarize, we have presented a model of $1/f$ flux noise in which electron
spins on the surface of metals relax via hyperfine interactions. Since the
electron spin relaxation time depends exponentially on the distance between
the electron and the nuclear spin, the nearest nuclear spin dominates the 
spin relaxation process. The distribution of distances results in a 
distribution $P(T_1^{\rm eff})$.
Averaging over this distribution results in $1/f$ flux noise. 
Experimentally, the SQUIDs producing flux noise are in steady state
equilibrium, so the noise is normally stationary and Gaussian. This is
what we have assumed in our calculations. (Stationary
means that the system, and hence the autocorrelation functions,
are translationally invariant in time. For Gaussian processes,
higher order correlation functions can be expressed as products
of the two-point (lowest order) correlation functions \cite{Kogan1996}.)
Since the magnetization sums over individual spins, the magnetization 
noise, and hence the flux noise, is Gaussian if there are enough
spins for the central limit theorem to apply. In both our model and experiment,
non-Gaussian noise could arise in very small samples \cite{Weissman1988}.

Our results indicate that
flux noise would be significantly reduced in superconducting materials
where the most abundant isotopes do not have nuclear moments such as zinc and lead.
The only isotopes of zinc and lead that have nuclear moments are $^{67}$Zn
and $^{207}$Pb which have natural isotopic abundances of 4\% and 22\%, respectively.  
Thus, compared to Nb SQUIDs, we would expect flux noise to be lower by 
roughly a factor of 25 and 5 in Zn and Pb SQUIDs since the relevant factor
is $\xi^2/\Delta r_n^2$ in Eq.~(\ref{eq:FluxS}). 
This is assuming that
Nb, Zn, and Pb have approximately the same atomic arrangement on their
surface with approximately the same density of surface spins.
For experimentally relevant values, the flux noise expression in 
Eq.~(\ref{eq:FluxS}) does not have any temperature dependence 
($\rm sech(\hbar\omega_0/k_B T)\approx 1$ since $\hbar\omega_0\ll k_B T$).
This is consistent with a quantum process such as hyperfine exchange coupling,
and with the plateau 
seen below 0.5 K in plots of the flux noise versus temperature \cite{Wellstood1987a}.
The unusual temperature dependence of the flux noise that is experimentally found above 
0.5 K \cite{Wellstood1987a} may involve thermal fluctuations of the spins.

We wish to thank Michael Weissman, Kam Moler and Hendrik Bluhm
for helpful discussions and suggestions. 
CCY thanks the Aspen Center for Physics (supported by NSF Grant
1066293) for their hospitality during which this work was initiated.
This work was supported by Army Research Office grant W911NF-10-1-0494.

\end{document}